\documentclass[traditabstract]{aa} 
\usepackage{graphicx}
\usepackage{txfonts}
\begin{document}
   \title{SDSS J162520.29+120308.7 - a new SU UMa star in the period gap}

   \author{A. Olech\inst{1}, E. de Miguel\inst{2,3}, M. Otulakowska\inst{1},
J.R. Thorstensen\inst{4},
A. Rutkowski\inst{5}, R. Novak\inst{6}, G. Masi\inst{7}, M. Richmond\inst{8},
B. Staels\inst{9,10}, S. Lowther\inst{11}, W. Stein\inst{9,12}, T. Ak\inst{5,13}, 
D. Boyd\inst{14,15}, R. Koff\inst{16}, J. Patterson\inst{17} 
          \and
          Z. Eker\inst{5}
          }

   \institute{Nicolaus Copernicus Astronomical Center, Bartycka 18, 00-176
              Warszawa, Poland\\
              \email{olech@camk.edu.pl}
	\and
Departamento de Fisica Aplicada, 
Facultad de Ciencias Experimentales, Universidad de Huelva, 21071 Huelva, Spain
	\and
Center for Backyard Astrophysics, Observatorio del CIECEM, Parque Dunar, 
Matalascanas, 21760 Almonte, Huelva, Spain
	\and
Dept. of Physics and Astronomy, 6127 Wilder Laboratory, Dartmouth College, Hanover, 
NH 03755-3528, USA
	\and
TUBITAK National Observatory, Akdeniz University Campus, 07058 Antalya, Turkey
         \and
Institute of Computer Science, Faculty of Civil Engineering, Brno University of 
Technology, 602 00 Brno, Czech Republic
	\and
The Virtual Telescope Project, Via Madonna del Loco 47, 03023 Ceccano (FR), Italy
	\and
Physics Department, Rochester Institute of Technology, Rochester, New York 14623, USA
	\and
American Association of Variable Star Observers, 49 Bay State Rd., Cambridge, 
MA 02138, USA
	\and
Center for Backyard Astrophysics (Flanders), American Association of Variable Star 
Observers (AAVSO), Alan Guth Observatory, Koningshofbaan 51, Hofstade, Aalst, Belgium
	\and
Center for Backyard Astrophysics (Pukekohe), New Zealand
	\and
6025 Calle Paraiso, Las Cruces, NM 88012, USA
	\and
Istanbul University, Faculty of Sciences, Department of Astronomy and Space Sciences,
34119 University, Istanbul, Turkey
	\and
Silver Lane, West Challow, Wantage, OX12 9TX, UK
	\and
The British Astronomical Association, Variable Star Section (BAA VSS), Burlington House, 
Piccadilly, London, W1J 0DU, UK
	\and
Center for Backyard Astrophysics (Colorado),
Antelope Hills Observatory, 980 Antelope Drive West,
Bennett, CO 80102, USA
	\and
Department of Astronomy, Columbia University, New York, NY 10027, USA
             }

   \date{Received January 11, 2011; accepted February 16, 2011}

\abstract{

We report results of an extensive world-wide observing campaign
devoted to the recently discovered dwarf nova SDSS J162520.29+120308.7 (SDSS
J1625). The data were obtained during the July 2010 eruption of the star and
in August and September 2010 when the object was in quiescence. During the
July 2010 superoutburst SDSS J1625 clearly displayed superhumps with a
mean period of $P_{\rm sh}=0.095942(17)$ days ($138.16 \pm 0.02$ min)
and a maximum amplitude reaching almost 0.4 mag. The superhump period was
not stable, decreasing very rapidly at a rate of $\dot P =
-1.63(14)\cdot 10^{-3}$ at the beginning of the superoutburst and
increasing at a rate of $\dot P = 2.81(20)\cdot 10^{-4}$ in the middle
phase. At the end of the superoutburst it stabilized around the value of
$P_{\rm sh}=0.09531(5)$ day.\\

During the first twelve hours of the superoutburst a low-amplitude
double wave modulation was observed whose properties are almost
identical to early superhumps observed in WZ Sge stars.  The period of
early superhumps, the period of modulations observed temporarily in
quiescence and the period derived from radial velocity variations are the
same within measurement errors, allowing us to estimate the most probable
orbital period of the binary to be $P_{\rm orb}=0.09111(15)$ days
($131.20 \pm 0.22$ min). This value clearly indicates that SDSS J1625 is
another dwarf nova in the period gap. Knowledge of the orbital and
superhump periods allows us to estimate the mass ratio of the system to
be $q\approx 0.25$. This high value poses serious problems both for the
thermal and tidal instability (TTI) model describing the behaviour of 
dwarf novae and for some models
explaining the origin of early superhumps.}

\authorrunning{Olech et al.}

\maketitle

\section{Introduction}

Dwarf novae are cataclysmic variables (CVs), close binary systems
containing a white dwarf primary and secondary star which fills its
Roche lobe. The secondary is typically a low mass main sequence star,
which loses its material through the inner Lagrangian point.  In the
presence of a weakly-magnetic white dwarf, this material forms an
accretion disc around the white dwarf (Warner 1995,  Hellier 2001).

One of the most intriguing classes of dwarf nova is the SU UMa type; its
members have short orbital periods (less than 2.5 hours) and show two
types of outbursts: normal outbursts and superoutbursts. Superoutbursts
are typically about one magnitude brighter than normal outbursts, occur
about ten times less frequently and display superhumps - characteristic
tooth-shaped light modulations with a period a few percent longer than
the orbital period of the binary.

The behaviour of SU UMa stars is not fully understood. Their peculiar
properties could be interpreted within the frame of the  thermal and
tidal instability (TTI) model (see Osaki 1996, 2005 for a review). 
According to this model superhumps are caused by prograde rotation of
the line of apsides of a disc elongated by tidal perturbation by the
secondary (very often and in fact incorrectly called ``precession'').
The perturbation is most effective when disc particles moving in
eccentric orbits enter the 3:1 resonance with the binary orbit. Then the
superhump period is simply the beat period between the orbital and
apparent disk precession periods (Whitehurst 1988, Hirose and Osaki
1990, Lubow 1991). However, this mechanism is effective in producing
superhumps only in the case of systems where the mass ratio $q$ is
smaller than 0.25 (Whitehurst 1988, Osaki 1989).

In recent years, several facts have appeared to argue against the TTI
model. A new and precise distance determination of SS Cyg shows that
both the accretion rate during outburst and the mean mass transfer rate
contradict the disc instability model (Schreiber and Lasota 2007).
Superoutbursts and superhumps are observed in systems with $q$
significantly larger than 0.25-0.3 (for example TU Men and U Gem - Smak
2006, Smak and Waagen 2004). There is clear evidence for the presence of
the hot spot during a superoutburst (Smak 2007, 2008) and indications
that the mass transfer rate during the eruption phase is 30-40 times
higher than in quiescence (Smak 2005). The amplitudes of bolometric
light curves produced by 2D and 3D smoothed-particle hydrodynamics (SPH)
simulations are about 10 times lower than real amplitudes of superhumps
observed in SU UMa systems (Smak 2009).  Finally, the multi-wavelength
predictions of enhanced mass transfer (EMT) model seem to fit
observations slightly better that predictions of TTI (Schreiber et al.
2004).

\begin{figure}
\centering
\includegraphics[width=9.5cm]{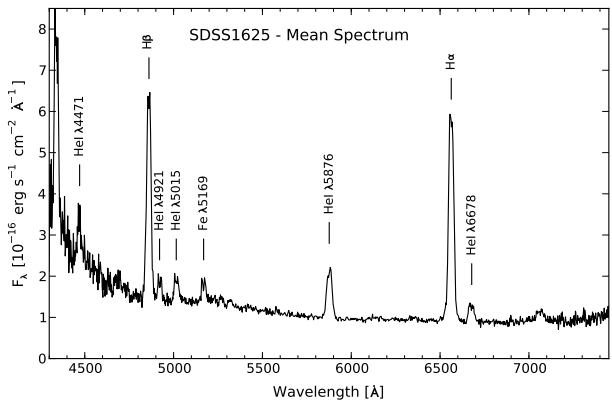}
\caption{The mean spectrum of SDSS J162520.29+120308.7 obtained in May 2010.}
\label{fig1}
\end{figure}

\begin{figure}
\centering
\includegraphics[width=9.5cm]{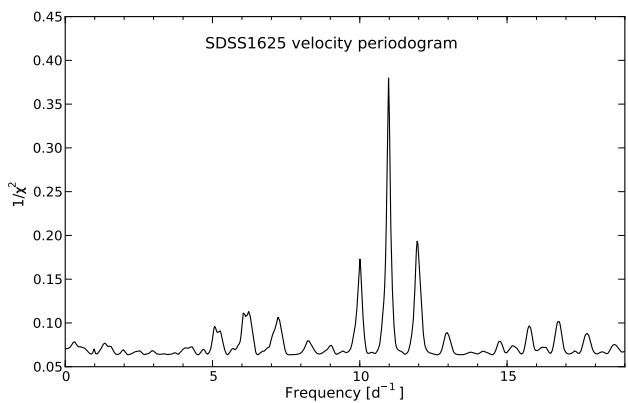}
\caption{The periodogram for the H$\alpha$ radial velocity data of 
SDSS J162520.29+120308.7.}
\label{fig2}
\end{figure}

There are also several unsolved problems in our knowledge of the
evolution of cataclysmic variable stars. A typical dwarf nova starts its
evolution as a binary system containing a $\sim0.6 {\cal M}_\odot$ white
dwarf and $0.2 - 0.5 {\cal M}_\odot$ main sequence secondary. Initially,
the orbital period is $\sim10$ hours and this decreases due to angular
momentum loss ($\dot J$) through magnetic braking via a magnetically
constrained stellar wind from the donor star (Hameury et al. 1988,
Howell et al. 1997, Kolb and  Baraffe 1999, Barker and Kolb 2003). At
this stage the mass-transfer rates are typically   $10^{-10} - 10^{-8}~
\cal{M}_\odot ~{\rm yr}$$^{-1}$. The CV evolves towards shorter periods
until the secondary becomes completely convective, 
at which point magnetic braking greatly decreases. 
This happens for an orbital period of around 3 hours.
Mass transfer significantly diminishes and the secondary shrinks towards
its equilibrium radius, well within the Roche lobe. Abrupt termination of
magnetic braking at $P_{\rm orb}\sim 3$ h produces a sharp gap,
between 2 and 3 hours, 
in the distribution of periods of dwarf novae. The binary reawakens as
a CV at $P_{\rm orb}\sim 2$ h when mass transfer recommences. It is
then driven mainly by angular momentum loss due to gravitational
radiation ($\dot J_{\rm GR}$). The orbital period continues to decrease while
mass transfer stays at an almost constant level.

It is clear that observations of systems located in the period gap are
very important. These systems challenge our understanding    of the
evolution of dwarf novae. In particular, their $q$ values are around
0.3-0.4, posing some problems for the classical TTI model (see for
example the latest determination of $q$ for TU Men in Smak 2006).

In this paper we present the results of a world-wide observational
campaign devoted to SDSS J162520.29+120308.7 - a newly discovered SU UMa
star with an orbital period within the period gap.

\begin{figure}
\centering
\includegraphics[width=9.5cm]{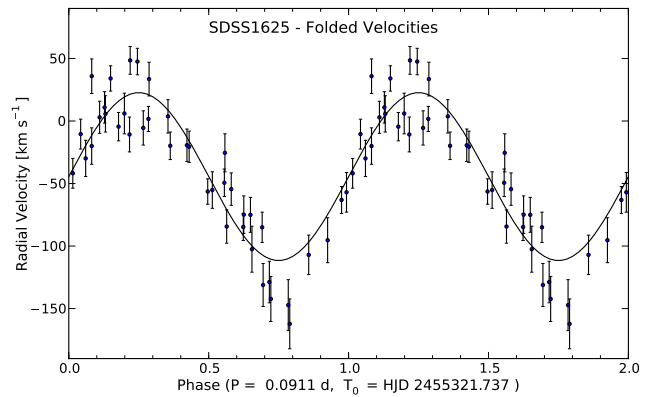}
\caption{The H$\alpha$ radial velocities of SDSS J162520.29+120308.7 
obtained in May 2010 folded on the best-fit period, with the
sinusoidal fit superposed.}
\label{fig3}
\end{figure}

\section{SDSS J162520.29+120308.7}

SDSS J162520.29+120308.7 (hereafter SDSS J1625) was identified as a
cataclysmic variable candidate by Wils et al. (2010). They pointed out
that the spectrum of the star is similar to that of RZ Leo and is
dominated by emission from the white dwarf and the M-dwarf companion
star, with broad emission lines superimposed.

\begin{table*}
\caption{The details of the observations collected during 2010 superoutburst
of SDSS J162520.29+120308.7}
\centering
\begin{tabular}{|l|r|l|r|r|r|}
\hline
Observer       &  Telescope &  Country   & No. of nights & Total time [h]& No. of frames \\
\hline
E. de Miguel   &  25 cm &  Spain      &16 & 52.395 &  3522\\
M. Otulakowska &  60 cm &  Poland     &13 & 44.845 &  1940\\
A. Rutkowski   & 100 cm &  Turkey     &10 & 25.802 &  1426\\
R. Novak       &  40 cm &  Czech Rep. & 7 & 29.670 &  2882\\
G. Masi        &  36 cm &  Italy      & 6 & 22.768 &  1247\\
M. Richmond    &  35 cm &  USA        & 5 & 19.914 &  1994\\
B. Staels      &  25 cm &  Belgium    & 4 & 14.187 &  1737\\
S. Lowther     &  25 cm &  New Zealand& 3 & 11.422 &   530\\
W. Stein       &  35 cm &  USA        & 2 &  8.914 &   886\\
T. Ak          &  40 cm &  Turkey     & 2 &  8.070 &   317\\
D. Boyd        &  35 cm &  United Kingdom & 2 &  6.266 &   363\\
R. Koff        &  25 cm &  USA        & 1 &  5.106 &   545\\
\hline
\end{tabular}
\end{table*}

Inspection of the Catalina Real-time Transient Survey (CRTS; Drake et
al. 2009) light curve led to the discovery of modulation of the
quiescent magnitude of the star with an amplitude of about 1 mag and
time-scale of around 500 days.

Recent photometric data for SDSS J1625 collected by CRTS show that the
star was at around 19.5 mag in 2007 and during the first half of 2008.
In the second part of 2008 and the whole year 2009 it was about 1 mag
brighter. In 2010 it brightened to around 18 mag.

SDSS J1625 was caught at the beginning of the outburst by CRTS in four
frames taken in the interval of 5:58 - 6:35 UT on 5 July 2010. The star
reached a $V$ magnitude of 13.10. About twelve hours later the first
regular runs of our photometric campaign started.

\begin{figure*}
\centering
\includegraphics[width=17.4cm]{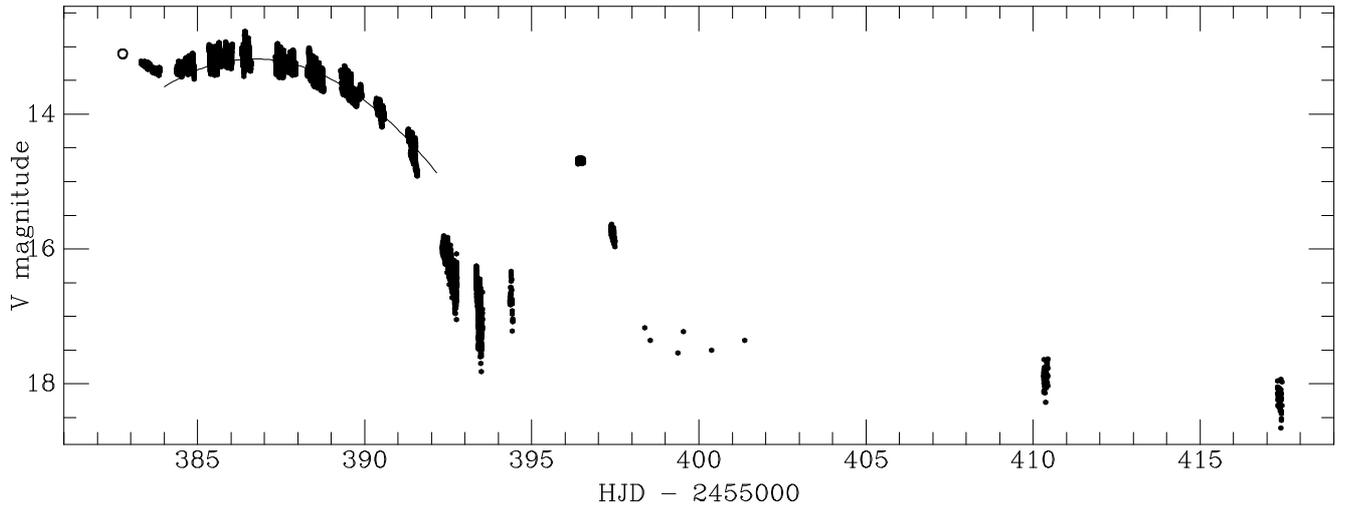}
\caption{Global light curve of the 2010 superoutburst of SDSS J162520.29+120308.7.
The solid line corresponds to the parabola fit to the 'plateau' phase.}
\label{fig4}
\end{figure*}

\section{Spectroscopy and Radial Velocities}

We obtained pre-outburst time-series spectra in 2010 May, using the
2.4-m Hiltner telescope at MDM Observatory on Kitt Peak, Arizona, USA.
The instrumentation, observing protocols, data reduction, and analysis
were as described by Thorstensen et al.~(2010).

The mean spectrum (Fig. 1) appears similar to the SDSS spectrum shown by
Wils et al. (2010).  The emission lines are all typical of dwarf novae
at minimum light.  H$\alpha$ has an emission equivalent width of $\sim
180$ \AA , and a FWHM of nearly 1400 km s$^{-1}$.  The weaker HeI lines
are notably double-peaked, with the peaks separated by $\sim 1000$ km
s$^{-1}$. At wavelengths $\lambda > 7000$ \AA , one starts to see a hint
of the contribution from the M dwarf; this is much more convincingly
detected in the SDSS spectrum. The sharp upturn we see at $\lambda <
4700$ \AA is almost certainly not as large as it appears; this is near
the end of our spectral range and flux calibration is uncertain in this
region.  The continuum flux suggests that the $V$ magnitude averaged
18.7 for our observations; experience suggests that this should be good
to $\pm 0.2$ mag or so.

The convolution function used to measure radial velocities of H$\alpha$
(see Schneider and Young, 1980 and Thorstensen et al. 2010 for details)
consisted of positive and negative Gaussian functions, each with a FWHM
of 275 km s$^{-1}$, separated by 1830 km s$^{-1}$. This weighted the
steep sides of the line profile, outside the line core.

Fig. 2 shows the results of a period search of the resulting
time series, computed using the `residual-gram' method described
by Thorstensen et al. (1996).  Because our velocities span a
range of nearly 7 h of hour angle, the daily cycle count is
unambiguously determined, and a frequency near 11 cycle d$^{-1}$
is selected.  The best-fit sinusoid of the form $v(t) = \gamma + K
\sin [ (2 \pi (t - T_0) / P]$ has
$T_0 = \hbox{HJD}~2455321.7371(17), P_{\rm spec} =  0.09111(15)\hbox{ day}, 
~K = 67(8)\hbox{ km s}^{-1}\hbox{, and }\gamma = -44(5)\hbox{ km s}^{-1}$.  
Fig. 3 shows the velocities folded on this best-fit period, with the   
sinusoidal fit superposed.

\begin{figure*}
\centering
\includegraphics[width=13.1cm]{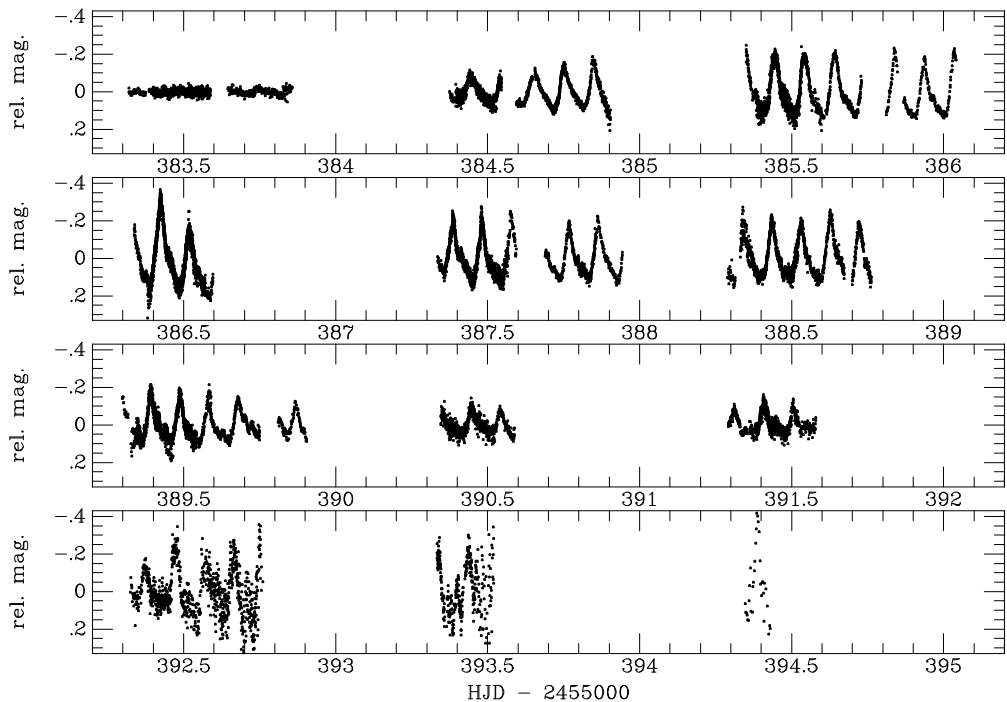}
\caption{Superhumps observed in 2010 superoutburst of
SDSS J162520.29+120308.7}
\label{fig5}
\end{figure*}

\section{Photometric Observations}

Our observing campaign started on 5 July 2010 and ended on 11 September
2010. In total, SDSS J1625 was observed in 71 runs by 12 observers using
12 telescopes with main mirror diameters ranging from 0.25 to 1.0 meter,
located in Europe, USA and New Zealand, operated mostly by  members of
the Center for Backyards Astrophysics (Patterson et al. 2003, 2005) and
the CURVE team (Olech et al 2008, 2009). The details are given in Table
1.

The star was observed for almost 250 hours and 17389 frames were
collected. "White light" or clear filters were used. The raw images
were flat-fielded and dark-subtracted by the observers. For
smaller telescopes the photometry was done using commercially available
software. For bigger telescopes IRAF routines\footnote{IRAF is
distributed by the National Optical Astronomy Observatory, which is
operated by the Association of Universities for Research in Astronomy,
Inc., under a cooperative agreement with the National Science
Foundation.} were used and profile-fitting photometry was derived using the
DaoPHOT/Allstar packages (Stetson 1987).

\section{Global light curve}

Fig. 4 shows the global light curve of SDSS J1625 in July 2010, and the
beginning of August 2010. The open circle corresponds to the averaged
observations of CRTS and shows that the star erupted with an amplitude of
$\sim 5$ mag.  After reaching a brightness of 13.1 mag at (truncated)
HJD 382.76, the star started to fade with a linear trend of around 0.25
mag/day, clearly indicating that we had observed an ordinary outburst (a
so-called precursor outburst), which then triggered the superoutburst. A
minimum at 13.4 mag was observed near HJD 384.0, after which SDSS J1625
began to increase its brightness going into superoutburst.

A typical SU UMa star, after a rapid initial increase in brightness,
enters the so-called 'plateau' phase in which its brightness shows a
roughly linear decreasing trend with a rate of $\sim 0.1 - 0.15$ mag per
day.  This phase usually lasts around 10-15 days. Some systems
additionally show a small rebrightening towards the end of the plateau
(see the  comprehensive discussion in Kato et al. 2003 and nice example
of rebrightening observed in TT Boo - Olech et al. 2004). SDSS
J1625 showed a slightly different behaviour, as can be seen in Fig. 4. The
plateau phase was relatively short, lasting slightly less than 8 days,
and can be described by a parabola (shown as a solid line) rather than
by a linear decreasing trend. It is interesting that similar behaviour
was recently observed in another SU UMa star in the period gap - SDSS
J162718.39+120435.0 (Shears et al. 2009).

Around HJD 392 the star entered the final decline stage and within two
days it dimmed to around 17 mag. It stayed at this level for another two
days, then went into a so-called echo outburst in which it reached
14.8 mag. The echo outburst finished around HJD 398 and for the
following four days the star stayed at a brightness of around 17.5 mag.
Further observations performed in August and September 2010 found SDSS
J1625 at a brightness below 18 mag.

\section{Superhumps}

Fig. 5 shows the evolution of the superhumps observed during the 2010
superoutburst of SDSS J1625. At the very beginning a low-amplitude 
($\sim 0.02$ mag) modulation with a period of around 0.045 days was
seen. A few hours later we see the birth of ordinary superhumps. Around
HJD 383.7 the first hump with a period of approximately 0.1 day and
amplitude of 0.035 mag is observed. During the next two days the
peak-to-peak superhump amplitude increases to almost 0.4 mag and the
light modulations gain their characteristic shark-tooth shape.

Around HJD 387, the superhump amplitude starts to decrease slowly but
shape remains constant for the next two days. Close to HJD 389, the
superhumps start to modify their shape, showing a weak secondary hump
near minimum light. These secondary humps never become strong, which is
typical of systems with longer orbital periods (see Rutkowski et al
2007). Near HJD 391, the amplitude of the light variations reaches a
minimum level of around 0.15 mag.

The evolution of the amplitude of superhumps is compared with 
that of the global light
curve of the 2010 superoutburst in the two upper panels of Fig.
6.

\subsection{O-C analysis}

To check the value and stability of the superhump period, we performed an
$O-C$ analysis for the maxima of the superhumps detected in the
superoutburst light curve. In total, we determined 46 times of maxima,
which are listed in Table 2 together with their cycle numbers, $E$, and
$O-C$ values.

\begin{table}
\caption{Cycle numbers, times of maxima and $O-C$ values for
superhumps observed in the 2010 superoutburst of SDSS J162520.29+120308.7.
The $E=0$ epoch corresponds to the moment of highest amplitude of
superhumps.}
\centering
\begin{tabular}{|r|c|r||r|c|r|}
\hline
$E$ & ${\rm HJD_{\rm max}}$ & $O-C$ & $E$ & ${\rm HJD_{\rm max}}$ & $O-C$\\
    & $ - 2455000$ & [day] & & $ - 2455000$ & [day] \\
\hline
$-$36 & 83.3428 & $-$0.893 & 17 & 88.5329 & 0.064\\
$-$35 & 83.4344 & $-$0.941 & 18 & 88.6283 & 0.056\\
$-$32 & 83.7475 & $-$0.686 & 19 & 88.7210 & 0.019\\
$-$25 & 84.4450 & $-$0.435 & 25 & 89.2995 & 0.034\\
$-$24 & 84.5420 & $-$0.426 & 26 & 89.3915 & $-$0.010\\
$-$23 & 84.6535 & $-$0.267 & 27 & 89.4883 & $-$0.004\\
$-$22 & 84.7530 & $-$0.233 & 28 & 89.5840 & $-$0.009\\
$-$21 & 84.8470 & $-$0.255 & 29 & 89.6790 & $-$0.021\\
$-$15 & 85.4457 & $-$0.031 & 31 & 89.8673 & $-$0.064\\
$-$14 & 85.5418 & $-$0.032 & 37 & 90.4465 & $-$0.042\\
$-$13 & 85.6420 & 0.010 & 38 & 90.5403 & $-$0.067\\
$-$11 & 85.8380 & 0.047 & 46 & 91.3119 & $-$0.045\\
$-$10 & 85.9380 & 0.087 & 47 & 91.4086 & $-$0.040\\
$-$9 & 86.0350 & 0.095 & 48 & 91.5042 & $-$0.046\\
$-$5 & 86.4238 & 0.137 & 57 & 92.3739 & $-$0.005\\
$-$4 & 86.5185 & 0.122 & 58 & 92.4711 & 0.006\\
5 & 87.3860 & 0.140 & 59 & 92.5670 & 0.003\\
6 & 87.4798 & 0.116 & 60 & 92.6625 & $-$0.004\\
7 & 87.5755 & 0.111 & 67 & 93.3413 & 0.052\\
9 & 87.7682 & 0.114 & 68 & 93.4365 & 0.042\\
10 & 87.8621 & 0.090 & 78 & 94.3870 & $-$0.076\\
15 & 88.3400 & 0.058 & 100 & 96.4900 & $-$0.213\\
16 & 88.4340 & 0.036 & 110 & 97.4390 & $-$0.347\\
\hline
\end{tabular}
\end{table}

\begin{figure}
\centering
\includegraphics[width=9cm]{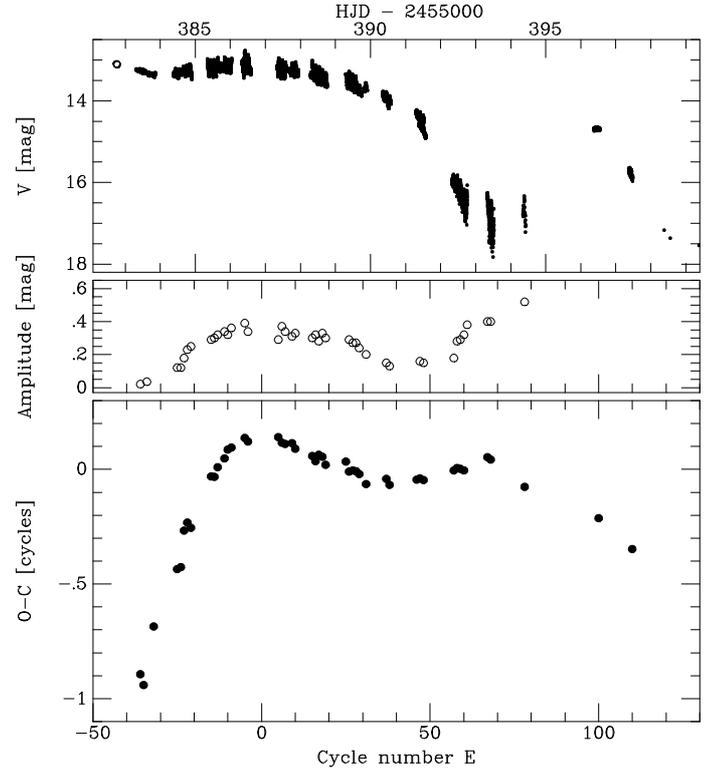}
\caption{Global light curve of the 2010 superoutburst of SDSS 
J162520.29+120308.7 (upper panel), evolution of the amplitude of 
superhumps (middle panel) and $O-C$ diagram (lower panel).}
\label{fig6}
\end{figure}

The cycle numbers $E$ and times of maxima were fitted with the
linear ephemeris:

\begin{equation}
{\rm HJD}_{\rm max} = 2455386.8915(3) + 0.096189(10) \times E
\end{equation}

\noindent indicating that the mean value of the superhump period
was $P_{\rm sh} = 0.096189$ day.

However, the $O-C$ diagram constructed using the above ephemeris (1) 
and shown in the lower panel of Fig. 6 clearly indicates the complex
pattern of superhump period changes. Considered globally, it is
consistent with the $ABC$ scenario recently suggested by Kato et al.
(2009). In phase $A$ (negative $E$ numbers), the period is significantly
longer than shown in (1); in phase $B$ ($0 < E < 70$), it is close to
the mean value but shows a clear increasing trend; and in phase $C$
($E>70$), the period seems to be constant and has a value shorter than
the mean value.

\begin{figure}
\centering
\includegraphics[width=9cm]{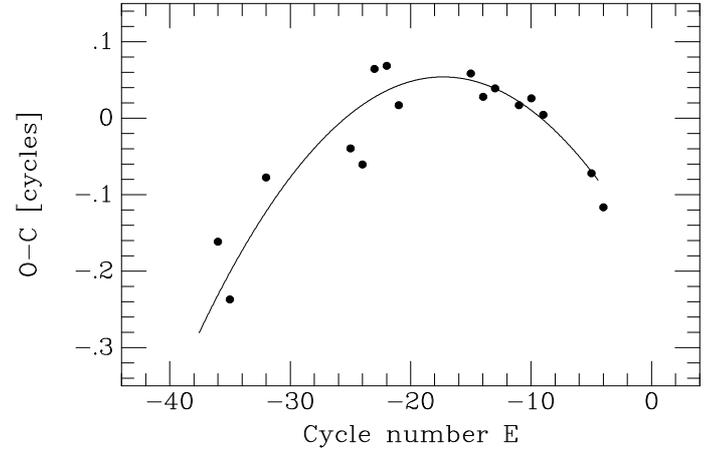}
\caption{$O-C$ diagram of phase A of the 2010 superoutburst of 
SDSS J162520.29+120308.7. The solid line corresponds to 
the ephemeris given by (3).}
\label{fig7}
\end{figure}

We decided to check the evolution of the superhump period 
by looking more closely
at its behaviour in each phase. First, we fitted the following linear
ephemeris to the negative $E$ numbers:

\begin{equation}
{\rm HJD}_{\rm max} = 2455386.9264(10) + 0.099102(67) \times E
\end{equation}

\noindent It is clear that the superhump period in this interval is
longer than the mean value shown in (1). To check its stability in phase
$A$, we plotted the $O-C$ diagram computed using (2).  The result is
shown in Fig. 7 and indicates that  the superhump period was decreasing
quickly at the beginning of the superoutburst. To obtain the $\dot P$
value, we constructed a quadratic ephemeris in the following form:

\begin{equation}
{\rm HJD}_{\rm max} = 2455386.9076(20) + 0.09630(26) \times E - \\
8.1(0.7) \cdot 10^{-5} \times E^2
\end{equation}

\noindent The resulting value of $\dot P$ equals to $-1.63(14)\cdot
10^{-3}$, which is very large - one of the highest ever observed in a SU
UMa system.

It is worth noting that the maxima at $E=-36$ and $E=-35$ are very
uncertain due to the small amplitude of the modulation observed at that
time. It is also possible that variations observed at the very beginning
of the superoutburst are not connected with ordinary superhumps (this
will be discussed in section 7). Omitting the first two points in the
$O-C$ diagram shown in Fig. 7 does not change our conclusion concerning
the very fast decrease of the superhump period during the early stage of
the superoutburst.

Let's move to phase $B$ of the superoutburst, i.e. the time interval
limited by the cycle values $E$ from 0 to 70. A linear fit to the times
of maxima determined in this interval is shown below:

\begin{equation}
{\rm HJD}_{\rm max} = 2455386.9006(6) + 0.095942(17) \times E
\end{equation}

\noindent There is a clear difference between the mean periods observed
in phases $A$ and $B$ amounting to 0.003 day, i.e. about 3\%.

From Fig. 7, we already know that in phase $B$ the superhump period was
increasing. The second order polynomial fit to the maxima occurring during
this stage is show in the following equation:

\begin{equation}
{\rm HJD}_{\rm max} = 2455386.9102(9) + 0.095067(66) \times E + \\
1.35(10) \cdot 10^{-5} \times E^2
\end{equation}

\noindent where the quadratic term corresponds to a $\dot P$ value of 
$2.81(20)\cdot 10^{-4}$.

Fig. 8 shows the $O-C$ diagram for phase $B$ of the superoutburst
constructed using (4).

\begin{figure}
\centering
\includegraphics[width=9cm]{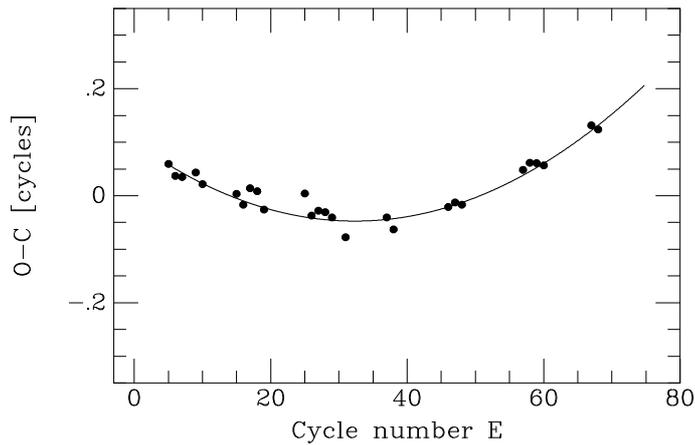}
\caption{$O-C$ diagram of phase B of the 2010 superoutburst of 
SDSS J162520.29+120308.7. The solid line corresponds to 
the ephemeris given by (5).}
\label{fig8}
\end{figure}

The amount of data collected at the end of the superoutburst (phase $C$)
does not allow us to draw any conclusions about the period derivative in
this interval, but the $O-C$ diagram in Fig. 6 suggests that the
superhump period in phase $C$ was constant. A linear fit to the times of
maxima with cycle numbers $E$ from 67 to 110 is shown below:

\begin{equation}
{\rm HJD}_{\rm max} = 2455386.9546(57) + 0.09531(5) \times E
\end{equation}

This indicates that, at the end of the superoutburst, the superhump
period was shorter than its mean value in phase $B$. This is consistent
with the scenario described recently by Kato et al. (2009).

\subsection{Power spectrum analysis}

The light curve from each individual run was fitted with a straight line
or parabola and this fit was then subtracted from the data. This was
done to remove the overall decreasing trend of the superoutburst and to bring
the mean value of all runs to zero. Next, the ANOVA power spectrum was
computed for the whole data set (Schwarzenberg-Czerny 1996). The
resulting spectrum is shown in the upper panel of Fig. 9. It shows a
clear peak at a frequency of $f=10.405(15)$ c/d with weak 1-day aliases.
This frequency corresponds to a period of 0.09611(14) day which is in
excellent agreement with the mean superhump period obtained in (1).

\begin{figure}
\centering
\includegraphics[width=9cm]{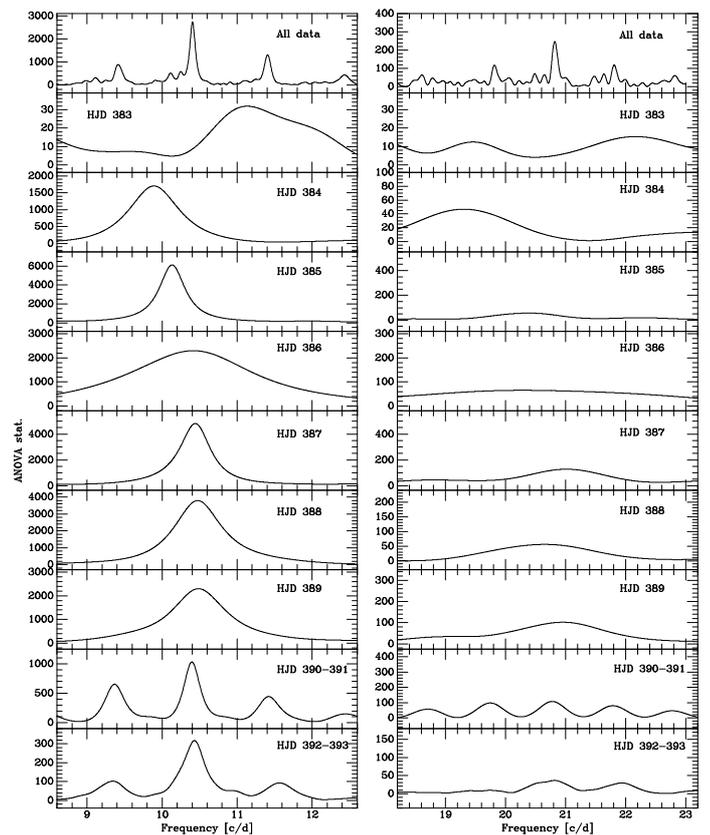}
\caption{ANOVA statistics for the whole data set and for 1-2 day blocks.}
\label{fig9}
\end{figure}

Knowing that the superhump period was changing during the superoutburst,
we decided to divide our data set into 1-2 day blocks and compute the
ANOVA power spectra for each of them separately. The large amount of
data collected in our campaign ensured that, even in the shortest block,
the number of data points amounted to 1281. The resulting power spectra
are shown in the corresponding panels of Fig. 9.

The power spectrum for HJD 383 is peculiar for two reasons. First, the
main peak is located far away from the strongest frequency obtained for
the whole data set. The most probable value of the frequency observed at
this very early stage of the superoutburst is $f=11.13(23)$ c/d,
corresponding to a period of 0.0898(18) day, which is significantly
shorter that the mean period of ordinary superhumps and within the error
close to the orbital period detected in spectroscopic data. The second
interesting point is that a strong peak is also observed at a frequency
of $2f$. This indicates that the modulation has a double wave structure
which is consistent with visual inspection of the light curve for this
interval.

Only one day later, at HJD 384, the situation changes completely. The
main peak is located at frequency $f=9.89(15)$ c/d and there is almost
no signal at $2f$. The corresponding superhump period is 0.1011(15) day
i.e. much longer than its mean value obtained for the whole data set.

As we already know from the $O-C$ data analysis, during phase $A$ the
superhump period was decreasing rapidly. This is fully confirmed by
analysis of the next data block. For HJD 385 the ANOVA spectrum shows a
very strong peak at frequency $f=10.133(130)$ c/d, again with only a
very weak signal at $2f$. This frequency corresponds to a period of
0.0987(13) day. This means that during one day the period shortened by
0.0024 day (3.5 minutes) from 0.1011 to 0.0987 day.

The ANOVA spectrum for the next block (HJD 386) suggests further period
decrease. The main peak is located at $f=10.42(27)$ c/d, which
corresponds to a period of 0.0960(25) day. 

According to the $O-C$ analysis, between HJD 386 and 387 the transition
from phase $A$ to $B$ was observed, at which point the superhump
amplitude reached a maximum of almost 0.4 mag.

For HJD 387 the ANOVA spectrum again shows a high peak, this time at a
frequency of $f=10.441(130)$ c/d corresponding to a period of 0.0958(12)
day. Within the uncertainties it is the same value as one day earlier.
The $O-C$ analysis performed for the superhump maxima observed in phase
$B$ showed a clear increase of the superhump period. However, the rate
of this change was an order of magnitude lower than in phase $A$. Thus,
it is expected that an increase in the period will not be visible in
one- day block periodograms. The frequency determination in such a short
interval suffers from an uncertainty of 0.1 c/d, corresponding to an
uncertainty in the period determination of $\sim 0.001$ day. Taking into
account the value of $\dot P$ determined for phase $B$, we can easily
calculate that during its six days the period increased by only 0.0017
day, a value comparable with the uncertainty in the period determination
for each one day interval. For this reason we abandon a detailed
description of the 1-2 day block periodograms and show only a short
summary of the period and power spectrum properties in Table 3.

\begin{table}
\caption{Values of frequencies and periods determined for the
different stages of the superoutburst.}
\centering
\begin{tabular}{|l|r|r|c|}
\hline
Block & $f$ [c/d]~~ & $P$ [days]~~ & $2f$ signal \\
\hline
All data &  10.405(15) & 0.09611(14) & medium \\
HJD 383  &  11.13(23)  & 0.0898(18) & very strong \\
HJD 384  &  9.89(15)   & 0.1011(15) & very weak \\
HJD 385  &  10.133(130) & 0.0987(13) & very weak \\
HJD 386  &  10.42(27)  & 0.0960(25) & weak \\
HJD 387  &  10.441(130) & 0.0958(12) & weak \\
HJD 388  &  10.475(150) & 0.0955(14) & weak \\
HJD 389  &  10.482(150) & 0.0954(14) & weak \\
HJD 390-391 & 10.397(90) & 0.0962(8) & medium \\
HJD 329-393 & 10.430(96) & 0.0959(8) & medium/weak \\
\hline
\end{tabular}
\end{table}

Finally, each data block was fitted with a Fourier sine series
containing from 6 to 10 harmonics and this fit was used for prewhitening
the original data. The ANOVA spectrum of the resulting prewhitened light
curve shows no significant peaks. Thus, we conclude that the superhump
signal was the one and only significant signal present during the
superoutburst.

\section{Quiescence}

The July 2010 superoutburst of SDSS J1625 ended around HJD 394 (July
16). Following this, the star went into a so-called echo eruption which
finished around HJD 398 (July 20). During the last 10 days of July, the
star remained close to its quiescent magnitude.

In August and September 2010, when SDSS J1625 was at $\sim 18.5$ mag, we
decided to perform two short campaigns whose main goal was to search for
orbital waves. On the nights of August 1/2, 2/3, 8/9, 11/12 and 12/13 we
observed the star using the 60-cm Cassegrain telescope in Ostrowik
Observatory, Poland. By this time the star was poorly placed and could
only be observed in the evening for 2.5 hours when it was low over the
western horizon (unfortunately polluted by the lights of the
neighbouring city). The resulting light curves are of poor quality (the
median value of the measurement error is 0.15 mag) and visual inspection
shows no clear modulation.

\begin{figure}
\centering
\includegraphics[width=9cm]{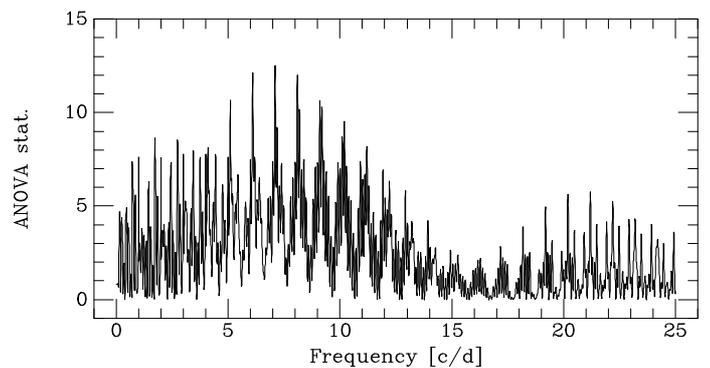}
\caption{ANOVA statistics for the data collected in the period of August 1/2 -
12/13 (HJD 410 - 421).}
\label{fig10} 
\end{figure}

In spite of the poor quality of the observations, we decided to compute the
ANOVA statistics for this data set. The resulting periodogram is shown
in Fig. 10. There are two distinct features in this plot: one group of
signals centered at frequency $f_1=7.096(23)$ c/d and another group
with the strongest signal at $f_2=21.196(23)$ c/d. This latter signal
is interesting. It suggests that the light curve consists
of a weak double humped wave with a period corresponding to $f_2/2$
i.e $P=0.09436(20)$ days. As the observing runs were short, the ANOVA
spectrum suffers significantly from aliases. Thus we cannot exclude the
possibility that the real period corresponds to one of the several aliases
located at frequencies between 20 and 23 c/d.

To find the correct value of the higher frequency, we 
prewhitened the data with a signal characterized by the frequency of
$f_1=7.096$ c/d. After doing this, the ANOVA power spectrum still shows
the highest peak at low frequencies - this time at 8.25 c/d. This second 
prewhitening results in a power spectrum showing a series of peaks in
a range between 22 and 27 c/d - all of them still above 3-sigma level. Due
to the aliasing problem, it is difficult to determine the correct value
of the frequency. It is interesting, however, that one peak is detected at
frequency of $f=22.11(3)$ c/d, suggesting the possibility of the presence
of double wave modulation with period of 0.0905(3) days. This value is
within 2-sigma consistent both with $P_{\rm spec}=0.09111(15)$ days
discussed in Sect. 3 and the period observed at very beginning of the
2010 superoutburst.

We are aware that the above analysis is speculative and based on the
poor quality data. However, there are two important things. First, the
power spectrum shown in Fig. 10 does not allow us to precisely determine
the appropriate frequencies but indicates that there are two signals in
the light curve: low frequency around 6-7 c/d and high frequency at
around 21 c/d. Second, the high frequency signal might be related
to double humped orbital wave.

The second campaign was performed between August 23/24 and September
11/12 when we observed SDSS J1625 on 8 nights using the 1.0-m R-C
telescope at TUBITAK National Observatory, Turkey. Example light curves
from four selected nights are shown in Fig. 11.

\begin{figure}
\centering
\includegraphics[width=8cm]{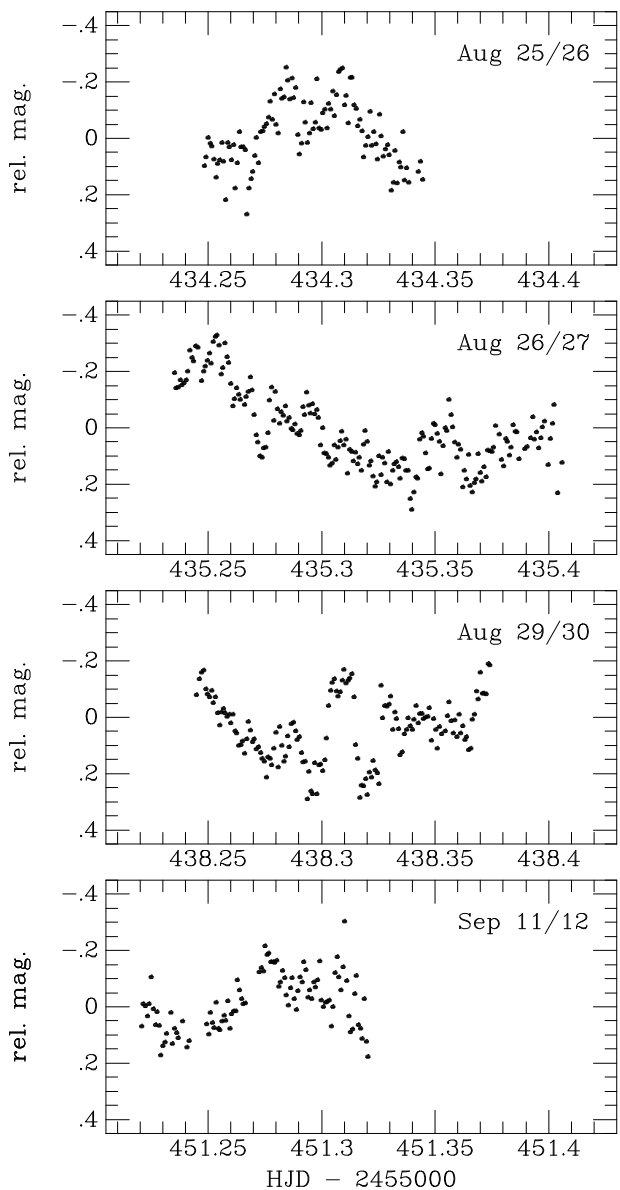}
\caption{Quiescent light curves of SDSS J1625 obtained in August
and September 2010 using a 1.0-m telescope.}
\label{fig11}
\end{figure}

\begin{figure}
\centering
\includegraphics[width=9cm]{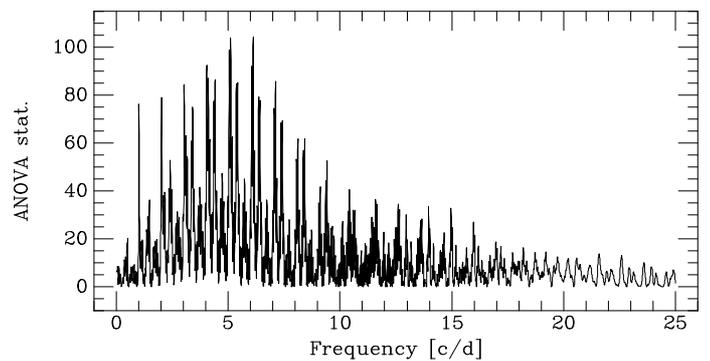}
\caption{ANOVA statistics for the data collected in the period August 23/24 -
September 11/12 (HJD 432 - 451).}
\label{fig12}
\end{figure}

As TUBITAK is a mountain observatory located 2.5 km above sea level, the
quality of the data was much better than for the Ostrowik runs. However,
the majority of the data was collected around Full Moon which occurred
on August 23.

Again we detrended the data and computed the ANOVA statistics which are
shown in Fig. 12. The spectrum is quite complicated but two distinct
signals (one real and one being its 1-day alias) with comparable power
are observed at frequencies of $f_1=5.112(15)$ and $f_2=6.115(15)$ c/d.
This low frequency modulation, corresponding to periods of 0.20 and 0.16
day respectively, is apparently visible in the light curves. It is most
distinct for the night of August 26/27, when its amplitude reaches over
0.3 mag. It is interesting SDSS J1625 is not the only object showing
such kind of variations (Rutkowski et al. 2010). The physical
interpretation of such a modulation is still lacking.

There is no evident signal which may be interpreted as the orbital
period of the binary. Prewhitening the original light curve with
frequencies $f_1$ or $f_2$ produces a light curve whose power spectrum
again shows some signal at frequencies around 5-6 c/d but no trace of
an orbital signal.

\section{Discussion}

In July 2010, SDSS J1625 experienced a $\sim 5$ mag eruption lasting
about 10 days. Detailed analysis of the light curve has enabled us to
detect clear short-period tooth-shaped modulations whose maximum
peak-to-peak amplitude reached almost 0.4 mag. These properties are
characteristic of SU UMa-type dwarf novae, indicating that SDSS J1625 is
a new member of this group.

The superhump period changed during the superoutburst, but as a
representative value we use the mean period for phase $B$ of the
superoutburst. Thus the mean superhump period of SDSS J1625 is $P_{\rm
sh} = 0.095942(17)$ day ($138.16\pm 0.02$ min). The orbital period of a
SU UMa star is typically a few percent shorter than the superhump period,
so we may expect that the orbital period of SDSS J1625 is around 130
minutes, i.e. inside the period gap.

Does our analysis reveal a candidate for the orbital period of the
binary? We believe so. First of all, we refer here to the phenomenon
called "early superhumps" observed in the early stages of superoutbursts
of WZ Sge type stars and first observed in the 1978 outburst of WZ Sge
itself by Patterson et al. (1981). The main properties of ``early
superhumps'' are:

\begin{itemize}

\item they are observed in the early stages of a superoutburst before
ordinary superhumps develop. In WZ Sge stars, this early stage may last
several days,

\item their period is very close to the orbital period of the binary. 
Ishioka et al. (2002) showed that the exact period of ``early superhumps'' is
only 0.05\% shorter that the orbital period of the binary,

\item they are double-wave modulations producing a power spectrum with
the strongest peak at the second harmonic.

\end{itemize}

There have been several attempts at a physical interpretation of ``early
superhumps.'' Patterson et al. (1981) suggested a ``super-hot spot'' model
in which early humps are due to a brightened hot spot, which in turn is
due to enhanced mass transfer from the secondary star during the
outburst. Kato et al. (1996) favor an ``early superhump'' interpretation
in which early humps are a premature form of true superhumps. Osaki
and Meyer (2002) introduce a model where ``early superhumps'' are the
manifestation of a tidal 2:1 resonance in the accretion disks of binary
systems with extremely low mass ratios. A similar model based mainly
on tidal distortion of a steady flow was suggested by Kato (2002).

There is an obvious candidate for ``early superhumps'' in the light
curve of SDSS J1625. It is the double-wave modulation observed in the
very early stage of the superoutburst (HJD 383), with a period
determined to be 0.0898(18) day. The only significant difference between
these humps and the ``early superhumps'' observed in WZ Sge systems is
their duration - hours in SDSS J1625 versus days in WZ Sge stars. This
property might be the reason why ``early superhumps'' were overlooked in
ordinary SU UMa stars and observed only in low $q$ systems where they
persist for much longer.

Could the period of 0.0898(18) day be associated with the orbital period
of the system? It is interesting that it is consistent within the
measurement errors with the modulation of 0.0904(3) day detected in the
quiescent light curve of the star collected in Ostrowik and  the period
of 0.09111(15) day determined from spectroscopy. 

There is one more test which can be done to check this hypothesis.
Cataclysmic variable stars showing superhumps are known to follow the
Stolz-Schoembs relation between the period excess $\epsilon$, defined as
$P_{\rm sh}/P_{\rm orb}-1$, and the orbital period of the binary (Stoltz
and Schoembs 1984). Fig. 13 shows this relation for over 100 dwarf novae
of different types (see the figure caption for a detailed description). 

Assuming that the orbital and superhump periods of SDSS J1625 are
$P_{\rm orb}=0.09111(15)$ and $P_{\rm sh}=0.095942(17)$ day
respectively, we calculate the period excess as $\epsilon = 0.053 \pm
0.002$. In Fig. 13, SDSS J1625 is plotted with a solid triangle. It is
located slightly above the mean linear fit but still within the global
trend, thus supporting our identification of the orbital period of the
binary. Recently, an even higher value of the period excess was noted in
V344 Lyr, which was observed by the Kepler mission (Still et al. 2010).

\begin{figure}
\centering
\includegraphics[width=9cm]{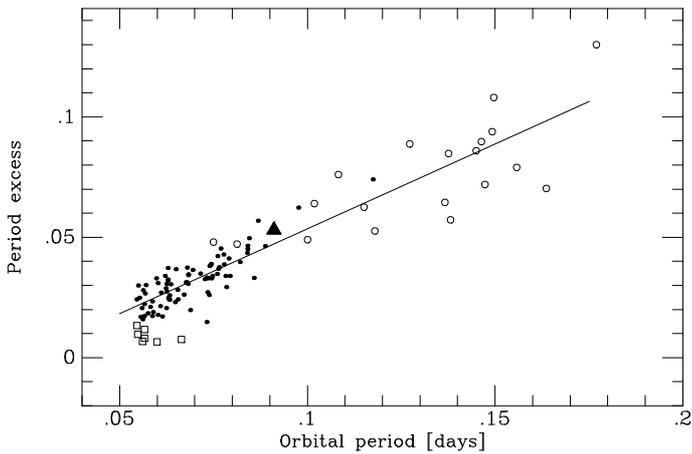}
\caption{Dependence between period excess and orbital period of
the binary for different types of cataclysmic variable. Ordinary SU UMa
stars are shown with dots. Open circles represent nova-like variables and
classical novae. The candidate period bouncers are plotted as open
squares. The position of SDSS J1625 corresponding to the adopted orbital 
period of 0.09111 days is marked with the solid triangle.}
\label{fig13}
\end{figure}

Knowing the period excess and using the empirical formula given by Patterson
(1998):

\begin{equation}
\epsilon\approx\frac{0.23q}{1+0.27q}
\end{equation}

\noindent we can estimate the mass ratio $q$. In the case of SDSS J1625
it is $q\approx 0.25$. Such a high mass ratio poses a problem not only
for the TTI model itself but for the explanation of ``early superhumps''
proposed by Osaki and Meyer (2002). Their model assumes a very low $q$
value which is needed to have sufficient matter in the disc to reach the
2:1 resonance radius. This condition is fulfilled only for $q<0.09$. For
systems with $q\approx 0.25$, the outer edge of the disc not only does not
reach the 2:1 resonance region but also has problems with accumulating 
the amount of matter at the 3:1 resonance region which, in the TTI model,
is needed to produce superhumps.

Moreover, up to now the highest $q$ system which showed ``early
superhumps'' was RZ Leo (Ishioka et al. 2001), a WZ Sge type star for
which $q$ is around 0.14.

Taking into account all the above facts, maybe we should not abandon the
original ``super hot-spot'' model proposed by Patterson et al. (1981), in
which early humps are due to a brightened hot spot which in turn is due
to enhanced mass transfer from the secondary. This is fully justified
by the recent work of Smak (2007, 2008), which shows clear evidence for the
presence of the hot spot during a superoutburst and indicates an
enhanced mass transfer rate during the eruption phase (Smak 2005).

One more exceptional phenomenon observed in SDSS J1625 is the very rapid
decrease of the superhump period during the early stage of the
superoutburst. Looking at the model light curves of superourbursts with
precursor computed both for TTI and EMT models by Schreiber et al.
(2004), one can see that this early phase of superoutburst is connected
with a decrease in the disc radius.  In the TTI model, the disc radius
decreases abruptly at the beginning and much slower during the rest of
``plateau'' phase. This abrupt decrease in the disc radius might be the
explanation of the very high $\dot P$ value observed during phase $A$.
In the EMT model, the decrease in disc radius  at the beginning of the
superoutburst is slower than in the TTI model, but still faster than in
the rest of the ``plateau'' phase. The problem is  that in both models,
during phase $B$, the radius of the disc still decreases but the period
of the superhumps increases.

\section{Summary}

\begin{enumerate}

\item Spectroscopic observations of SDSS J1625 performed in  quiescence
showed radial velocity modulations with a period of $P_{\rm
spec}=0.09111(15)$ day.

\item In July 2010, SDSS J1625 experienced a superoutburst lasting for
about 10 days and having an amplitude of around 5 mag. After the end of
the eruption, one echo outburst was observed and the star then returned
to its quiescent magnitude.

\item Superhumps were observed during almost the entire duration of the
superoutburst. Their mean period in the middle phase of the
superoutburst was $P_{\rm sh}=0.095942(17)$ day ($138.16 \pm 0.02$ min)
and their amplitude reached almost 0.4 mag.

\item The superhump period was not stable, decreasing very rapidly at a
rate of $\dot P = -1.63(14)\cdot 10^{-3}$ at the beginning of the
superoutburst and increasing at a rate of $\dot P = 2.81(20)\cdot
10^{-4}$ in the middle phase. At the end of the superoutburst it
stabilized around the value of $P_{\rm sh}=0.09531(5)$ day.

\item During the first twelve hours of the superoutburst a low-amplitude,
double-wave modulation was observed whose properties are almost
identical to the ``early superhumps'' observed in WZ Sge stars. 

\item The period of the ``early superhumps,'' the period determined from
spectroscopy, and the period of modulations observed temporarily in
quiescence are the same within measurement errors, allowing us to
estimate the most probable orbital period of the binary to be  $P_{\rm
orb}=0.09111(15)$ day ($131.20 \pm 0.22$ min). This value indicates that
SDSS J1625 is another dwarf nova in the period gap. Except of SDSS
J1625, there are only eleven known SU UMa stars with precisely determined
orbital periods with values between 2 and 3 hours.

\item From the orbital and superhump periods, we find a period excess
$\epsilon = 0.053 \pm 0.002$, which in turn provides a mass ratio
estimate of $q\approx 0.25$.

\item Such a high value of mass ratio may pose problem for some models
(eg. Osaki and Meyer 2002) which explain early superhumps, and may
support the idea of enhanced mass transfer and increased brightness of
the hot spot during the early phase of the superoutburst.

\item Quiescent light curves of SDSS J1625 seem to consist of a lower
frequency signal with strongest peaks in the range of 5-8 c/d. There
is still no explanation for this kind of modulation.

\end{enumerate}

\begin{acknowledgements}
This work was supported by the MNiSzW grant no. N203~301~335 to A. Olech.
A. Rutkowski has been supported by 2221-Visiting Scientist Fellowship
Program of TUBITAK. We thank to TUBITAK for a partial support in using 
T100 and T40 telescopes with project number 10CT100-94. JRT acknowledges 
support from the U.S. National Science Foundation, through grant AST 0708810.
\end{acknowledgements}

\end{document}